\documentstyle[12pt,psfig]{article}

\begin{document}

\newcommand{\ep}{equivalence principle~}
\newcommand{\Ch}{Chandrasekhar~}
\newcommand{\Chp}{Chandrasekhar}
\newcommand{\Sc}{Schwarzschild~}
\newcommand{\Scp}{Schwarzschild}
\newcommand{\Sw}{Schwarzschild~}
\newcommand{\Swp}{Schwarzschild}
\newcommand{\Sch}{Schr{\"{o}}dinger~}
\newcommand{\Schp}{Schr{\"{o}}dinger}
\newcommand{\OVp}{Oppenheimer--Volkoff}
\newcommand{\OV}{Oppenheimer--Volkoff~}
\newcommand{\GR}{General Relativity~}
\newcommand{\GT}{General Theory of Relativity~}
\newcommand{\GRp}{General Relativity}
\newcommand{\GTp}{General Theory of Relativity}
\newcommand{\STp}{Special Theory of Relativity}
\newcommand{\ST}{Special Theory of Relativity~}
\newcommand{\Lt}{Lorentz transformation~}
\newcommand{\Ltp}{Lorentz transformation}
\newcommand{\rel}{relativistic~}
\newcommand{\relp}{relativistic}
\newcommand{\msun}{M_{\odot}}
\newcommand{\eos}{equation of state~}
\newcommand{\eoss}{equations of state~}
\newcommand{\eossp}{equations of state}
\newcommand{\eosp}{equation of state}
\newcommand{\Eos}{Equation of state}
\newcommand{\Eosp}{Equation of state}
\newcommand{\beqn}{\begin{eqnarray}}
\newcommand{\eeqn}{\end{eqnarray}}
\newcommand{\nonum}{\nonumber \\}
\newcommand{\walecka}{$\sigma,\omega,\rho$~}
\newcommand{\waleckap}{$\sigma,\omega,\rho$}
\newcommand{\bbar}[1] {\mbox{$\overline{#1}$}} 
\newcommand{\courtesy}{~Reprinted with permission of Springer--Verlag 
New York; copyright 1997}
\newcommand{\oo}{{\"{o}}}
\newcommand{\au}{{\"{a}}}


\newcommand{\approxlt} {\mbox {$\stackrel{{\textstyle<}} {_\sim}$}}
\newcommand{\approxgt} {\mbox {$\stackrel{{\textstyle>}} {_\sim}$}}
\newcommand{\mearth} {\mbox {$M_\oplus$}}
\newcommand{\rearth} {\mbox {$R_\oplus$}}
\newcommand{\eo} {\mbox{$\epsilon_0$}}
\newcommand{\bfalpha} {\mbox{\mbox{\boldmath$\alpha$}}}
\newcommand{\bfgamma} {\mbox{\mbox{\boldmath$\gamma$}}}
\newcommand{\bfrho} {\mbox{\mbox{\boldmath$\rho$}}}
\newcommand{\bfsigma} {\mbox{\mbox{\boldmath$\sigma$}}}
\newcommand{\bftau} {\mbox{\mbox{\boldmath$\tau$}}}
\newcommand{\bfLambda} {\mbox{\mbox{\boldmath$\Lambda$}}}
\newcommand{\bfpi} {\mbox{\mbox{\boldmath$\pi$}}}
\newcommand{\bfomega} {\mbox{\mbox{\boldmath$\omega$}}}
\newcommand{\bp} {\mbox{\mbox{\boldmath$p$}}}
\newcommand{\br} {\mbox{\mbox{\boldmath$r$}}}
\newcommand{\bx} {\mbox{\mbox{\boldmath$x$}}}
\newcommand{\bvee} {\mbox{\mbox{\boldmath$v$}}}
\newcommand{\bu} {\mbox{\mbox{\boldmath$u$}}}
\newcommand{\bk} {\mbox{\mbox{\boldmath$k$}}}
\newcommand{\bA} {\mbox{\mbox{\boldmath$A$}}}
\newcommand{\bB} {\mbox{\mbox{\boldmath$B$}}}
\newcommand{\bF} {\mbox{\mbox{\boldmath$F$}}}
\newcommand{\bI} {\mbox{\mbox{\boldmath$I$}}}
\newcommand{\bJ} {\mbox{\mbox{\boldmath$J$}}}
\newcommand{\bK} {\mbox{\mbox{\boldmath$K$}}}
\newcommand{\bP} {\mbox{\mbox{\boldmath$P$}}}
\newcommand{\bS} {\mbox{\mbox{\boldmath$S$}}}
\newcommand{\bdel} {\mbox{\mbox{\boldmath$\bigtriangledown$}}}

\newcommand{\eps} {\mbox {$\epsilon$}}
\newcommand{\gpercm} {\mbox {${\rm g}/{\rm cm}^{3}$}}
\newcommand{\rhon} {\mbox {$\rho_{0}$}}
\newcommand{\rhoc} {\mbox {$\rho_{c}$}}
\newcommand{\fmm} {\mbox {${\rm fm}^{-3}$}}
\newcommand{\bag} {\mbox {$B^{1/4}$}}

\newcommand{\fraca} {\mbox {$ \frac{1}{2}       $}}
\newcommand{\fracb} {\mbox {$ \frac{3}{2}       $}}
\newcommand{\fracc} {\mbox {$ \frac{1}{4}       $}}
\newcommand{\fraccc} {\mbox {$ \frac{1}{3}       $}}

\newcommand{\tit}
{Model-independent Mass-Radius Constraint for Neutron Stars}

\newcommand{\auth} {Norman K. Glendenning}
\newcommand{\lbl}{LBNL-44366}
\newcommand{\dateofdoc}{\today}
\newcommand{\adr} 
{
Nuclear Science Division \& 
Institute for Nuclear and Particle Astrophysics
  Lawrence Berkeley  National Laboratory,
   MS: 70A-3307 \\ Berkeley, California 94720}

\newcommand{\doe}
{This work was supported by the
Director, Office of Energy Research,
Office of High Energy
and Nuclear Physics,
Division of Nuclear Physics,
of the U.S. Department of Energy under Contract
DE-AC03-76SF00098.}

\newcommand{\ect}{A part of this work was done at the ECT*,
Villa Tambosi, Trento, Italy.}


\begin{titlepage}
\begin{center}
\parbox{3in}{\begin{flushleft}Preprint \end{flushleft}}%
\parbox{3in}{\begin{flushright} \lbl  \end{flushright}}
~\\[7ex]

\renewcommand{\thefootnote}{\fnsymbol{footnote}}
\setcounter{footnote}{1}

\begin{Large}
\tit {\footnote{\doe}}\\[2ex]
\end{Large}
\renewcommand{\thefootnote}{\fnsymbol{footnote}}
\setcounter{footnote}{1}
~~\footnotetext{\tiny{[nkg/papers/radius/radius.tex,  \today} }

\begin{large}
\auth\\[3ex]
\end{large}
\adr\\[2ex]
\dateofdoc \\[2ex]
\end{center}
\begin{figure}[htb]
\vspace{.4in}
\begin{center}
\leavevmode
\hspace{-.2in}
\psfig{figure=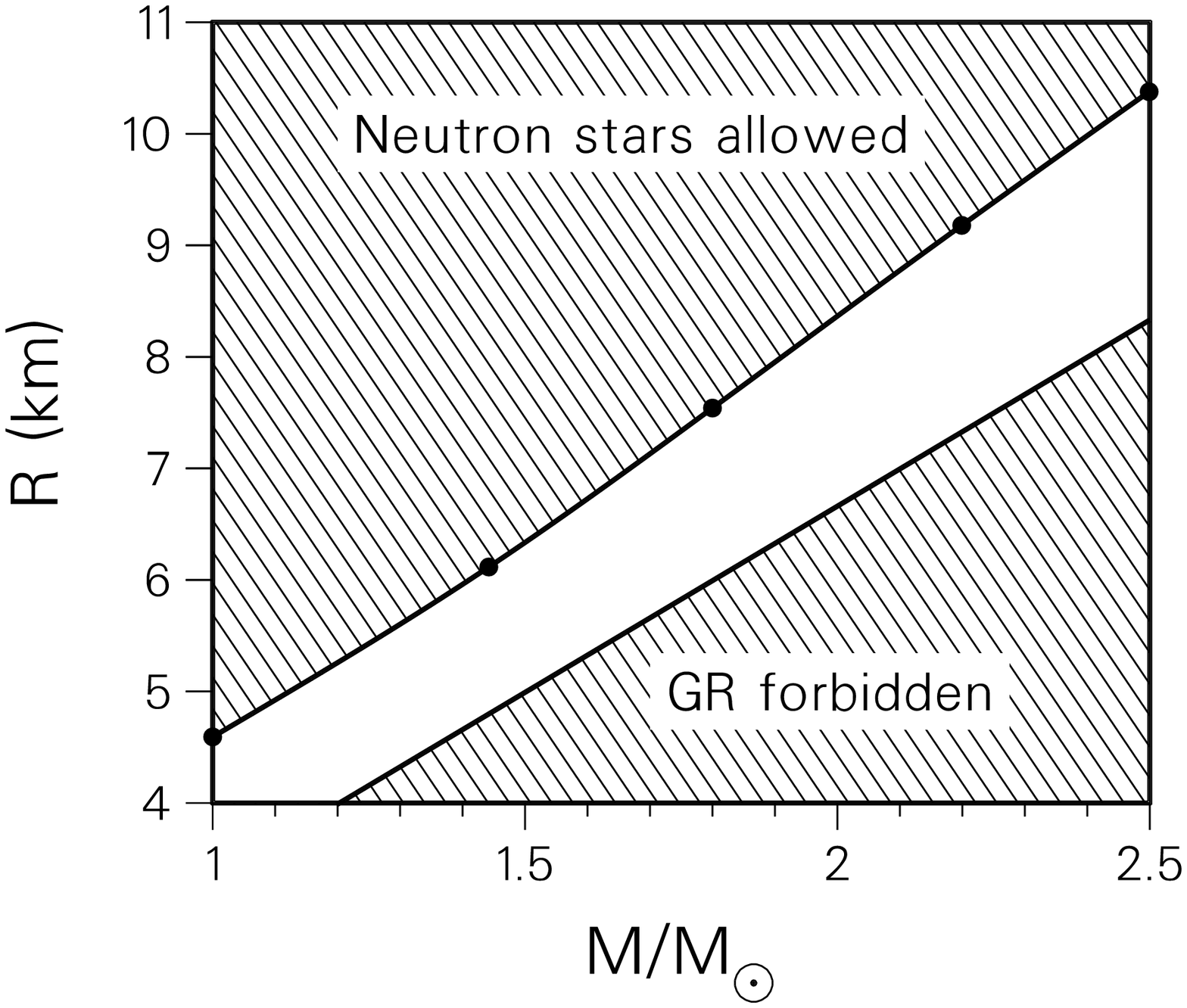,height=3.2in}
\end{center}
\end{figure}


\end{titlepage}

\clearpage

\setcounter{page}{0}
\newpage
~~~
\newpage


\begin{center}
\begin{Large}
\tit \\[7ex]
\end{Large}

\begin{large}
\auth \\[2ex]
\end{large}
\adr
\end{center}


   \begin{abstract}

While model-independent limits are always interesting, a limit on
neutron star
radius as a function of mass attains special interest
in light of recent interpretations of the periodic as well as
quasi-periodic oscillations
(QPOs)
in brightness of X-rays emitted from neutron stars that are accreting
matter from a low-mass companion. Here we derive such a limit based
only on well accepted principles.   We discuss our limit
 in connection with a recent
interpretation of X-ray pulsations  from SAX J1808.4-3658
as indicating a strange-star candidate, and show that this object
can also be a normal neutron star, though one whose central
core has  very high density.
 The most plausible high-density phase of hadronic
matter, which is also expected to be very compressible,
is quark matter. So an alternative to the strange star interpretation
of  SAX J1808.4-3658 is that it is  a hybrid
neutron star. 

\end{abstract}

%

\section{Motivation}

An accreting X-ray binary having millisecond
pulses was recently discovered (\cite{klis98:b,chakrabarty98:a}).
Presumably it is one of 
a class of objects that represent
the missing link between canonical  and millisecond
pulsars. If so,
it  confirms a long-standing conjecture that millisecond pulsars
are formed by accretion onto  canonical pulsars
(\cite{alpar82:a,bachus82:a,radhakrishnan82:a}). 
Many other
accreting X-ray sources have been discovered consisting
of a neutron star and a low-mass companion from which
material is gathered, presumably, 
from an accretion disk
(\cite{klis96:a,strohmayer96:a,klis98:a,klis99:a}).  An interpretation
of quasi-periodic oscillations in X-ray brightness 
characteristic of these sources suggests that
upper limits on  mass and
radius of the neutron star may  possibly be deduced
(\cite{strohmayer96:a,lamb98:c}).  
However, the interpretation in such terms is the subject of
some controversy and it remains to be seen how the models 
of the observations finally play out (\cite{wasserman99:a}). 
Certainly there is much uncertainty concerning the magnetic
field and accretion disk interaction which play an important role
in the modeling of X-ray pulsations.

Tentative limits on mass and radius deduced from models of
quasi periodic oscillations in X-ray brightness (QPOs)  
 have been  employed recently to 
discriminate among models of the equation of state of 
dense nuclear matter
(\cite{schaab99:a}). The X-ray pulsar, Sax J1808.4-3658,
is a particularly interesting object; it produces coherent X-ray
emission with a 2.5 ms period as well as X-ray bursts.
Based on an analysis of radiation from this object, a limiting 
mass-radius relationship
was derived which is difficult to reconcile with existing neutron star 
models  (\cite{heuvel99:a}). The mass-radius
relationship derived would be consistent with an interpretation
of Sax J1808.4
as
a strange star candidate as found by the above authors. 

Against this background
our purpose is
to derive a model-independent mass-radius
 constraint for neutron stars
that depends only
on minimal and well accepted principles. The limiting
relation is analogous to a previously obtained lower limit on the
Kepler period of a rotating star as a function of its mass
(\cite{glen92:a,friedman96:a}), and to an even earlier analysis
of limits on the gravitational redshift from neutron stars
(\cite{lindblom84:a}).

The most conservative minimal principles 
and constraints are:\\
1.  Einstein's general relativistic
equations for stellar structure hold.\\
2. The matter of the star
satisfies $dp/d\rho \geq 0$
which is a necessary condition
that a body is stable,
both as a whole and also with respect to the spontaneous
expansion or contraction of elementary regions away from equilibrium
(Le Chatelier's principle).\\
3. The \eos satisfies the
causal constraint for a perfect fluid;
a sound signal cannot propagate
faster than the speed of light,
$v(\epsilon)
\equiv\sqrt{dp/d\epsilon} \leq 1 $,
which is also the appropriate expression for sound signals
in General
Relativity (\cite{curtis50:a,geroch91:a}).\\
4.  The high-density
equation of state matches continuously
in energy and pressure to the
low-density equation of state of
\cite{baym71:b} and has no bound state at any density.

The last condition assures that the $M-R$ 
relation obtained  is for a neutron star
and not some sort of exotic. We mean ``neutron star'' in the generic
sense: it is made of charge neutral nuclear
matter at low density, while at higher density in the interior,
matter may be in a mixed or pure
quark-matter or other high-density phase of nuclear matter.
The last condition also  implies that the star is bound by gravity
as is a neutron star and is not a self-bound star such as a strange
star. As we will see, a self-bound star can lie in a region of the
$M-R$ plane that is forbidden to neutron stars.
In referring to the constraints as conservative, we mean that
we make no assumption about dense matter aside from the constraints mentioned
and we specifically allow for a phase transition above a baryon density
of $0.1625 {\rm~fm^{-3}}$. We discuss this further in the Section,
Caveats.

We can adapt the results of our earlier search
for a model-independent minimum Kepler period by searching for
the  radius at fixed mass that  minimizes
$P\sim (R^3/M)^{1/2}$ (\cite{glen92:a}). 
Several researchers found that the 
above classical result applies to relativistic stars to within a 
few percent accuracy  with a 
suitable constant of proportionality (\cite{haensel89:a,friedman89:a}).
We use variational \eoss subject to the above constraints and
techniques as described in the above reference. 
Our earlier results for 
the Kepler period agree to six percent 
with the results of 
 \cite{friedman96:a}, who performed a numerical
solution for rotating stars in place of the above approximation
formula for the Kepler period in terms of mass and radius of the
non-rotating counterparts.

\begin{figure}[htb]
\begin{center}
\leavevmode
\hspace{-.2in}   \psfig{figure=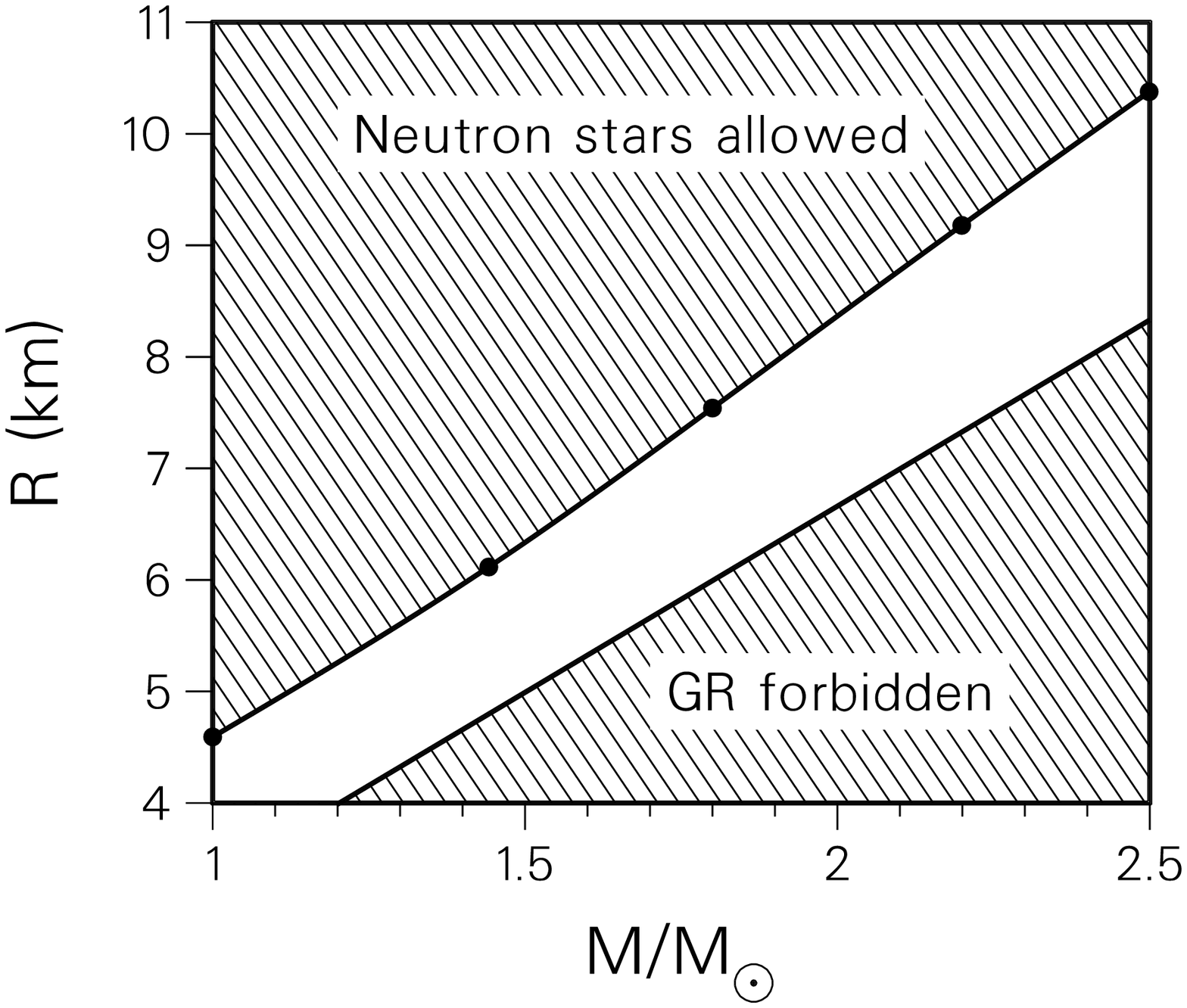,width=3in}
\parbox[t]{4.4 in}
{ \caption{ \label{rlimit} Allowed region of
non-rotating
neutron star radii as a function of limiting mass determined independent of
specific models of dense matter, but rather by the
minimal constraints (causality, etc) enumerated in the text.
The variational results are indicated by the dots. Stars that are not at the 
mass
limit will have larger radii.
}}
\end{center}
\end{figure}

Our results are shown in Fig.\ \ref{rlimit}. Neutron stars at the mass limit
can have radii as small as those shown by the line, and otherwise
must lie in the shaded region marked for neutron stars. The region
can be approximated in the interval illustrated  by 
\beqn
R\geq \Bigl( 3.1125 - 0.44192 x + 2.3089 x^2 - 0.38698 x^3 \Bigr)  {\rm~km},~~
\nonumber \\(1\leq x\equiv \frac{M}{\msun} \leq 2.5)\, .~~~~~~~
\eeqn
Of course, for neutron stars (unlike white dwarfs),
there is only one \eos in nature; all
neutron stars form a single family, and whatever the trajectory of the
mass-radius relationship is for that family,   the
limiting mass  star has the smallest radius,
and it is greater or equal to the limit derived.
For example, if the most massive neutron star that could exist in nature,
independent of formation mechanism, is $2 \msun$, the radius of all
neutron stars would have to exceed 8.37 km.
(Recall that measured masses tell us nothing about the maximum possible
mass that can be supported by nature's \eosp.)

Another limit of interest follows from the properties of
General Relativity. Schartzs\-child's limit $R>2M$ is 
actually less stringent than $R>9M/4$, which must be obeyed by 
any relativistic star (\cite{buchdahl59:a,weinberg}).
The latter is also plotted. If a star's mass
and radius placed it in the region
between the above described regions,
it could be made of matter that is self-bound at
high density, matter that would
be bound in microscopic to stellar like-objects even in the
absence of gravity (see Eq.\ \ref{den}). Strange
stars, if the strange matter hypothesis is true, are   examples. 

\section{Application to X-ray Emitters}

In the approximations and hypotheses that have been used to interpret 
the oscillations in X-ray luminosity, there appears a Keplerian
radius. Such a radius expresses the balance of gravitational
and centrifugal forces. In classical physics as well as in General
Relativity for a {\sl non-rotating} star
(units are
$G=c=1$):
\beqn
\Omega= \sqrt{M/R_K^3}\, , \label{kepler}
\eeqn
where $M$ is the mass of the star and 
 $\Omega$ is the  angular velocity of a particle 
in circular orbit at $R_K$.
This relation has nothing to do with the nature of the 
interior of the star,
whatever that may be. It
 only relates gravity in the exterior region to the centrifugal
force on a particle at $R_K$. 
The expression is exact for a {\sl non-rotating}
 star in General Relativity,
and only approximate for a rotating star, but there is a multiplicative
factor on the right side,
found to be accurate for many models that have been tested,
which is $\zeta\approx 0.65$
(\cite{haensel89:a,friedman89:a}).

The same relationship  $M/R^3=const$ holds also 
 for a self-bound object, such as a strange star as
 can be seen as follows: The average
energy density  $\bar{\epsilon}$
satisfies the identity 
\beqn
\epsilon_{{\rm equil.}}\leq
\bar{\epsilon} \equiv M \Big/ \Big(\frac{4\pi}{3}R^3 \Bigr)
\label{den}
\eeqn
where $\epsilon_{{\rm equil.}}$ denotes the equilibrium density
at which hypothetical strange matter is bound.
 The equality would hold
for spherical objects of mass such that gravity is unimportant.
Thus in either case, $M/R^3=const$ characterizes both  a Keplerian
orbit and a strange star, though they have nothing to do with
each other \`{a} priori.
 We make this point since superficially the  constraint derived from
 X-ray luminosity oscillations
looks like the $M-R$ relationship for a strange star
[cf.  Fig.\ 3 in Ref.\ (\cite{alcock86:a})].

\begin{figure}[htb]
\begin{center}
\leavevmode
\hspace{-.2in}   \psfig{figure=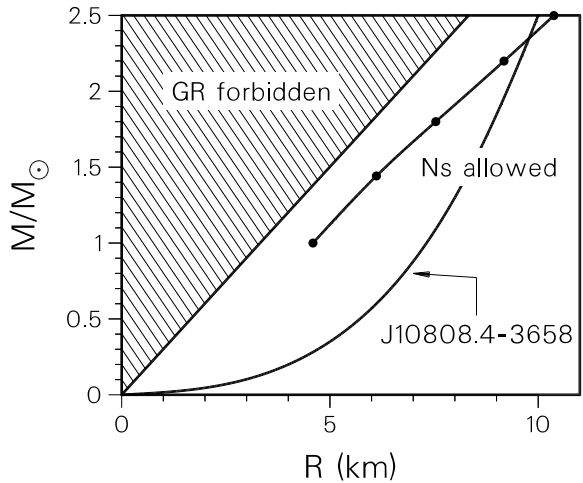,width=3in}
\parbox[t]{4.4 in} 
{ \caption { \label{slimit} According to the
QPO model of \protect\cite{heuvel99:a} the X-ray star must lie
on or to the left of the curved line $R\propto M^3$ and
by our model independent determination, neutron stars
must lie to the right of the
line marked `Ns allowed'. }}
\end{center}
\end{figure}

In fact, strange stars have been considered as 
candidates that satisfy the constraints of the  QPO 
model
(\cite{schaab99:a}) and of a model of
of periodic  pulsations from the X-ray pulsar
(\cite{heuvel99:a}). 
The suggestions {\sl seem} especially appealing because of the 
above coincidence. They also have some appeal in as much as many 
explicit neutron star models cannot satisfy the constraints
imposed by the theoretical analyses. Figure
\ref{slimit} illustrates the limit obtained for the $2.5$ ms 
X-ray pulsar, Sax J1808.4, by  Li et al. (1999), 
together with the limit obtained in our model independent way.
Neutron stars must lie to the right of our limit and the X-ray object
must lie on or to the left of the line so marked. Li et al. (1999)
have proposed the object as a strange-star candidate because
the neutron star models they tested did not meet their constraint, whereas
the strange star models did. However, from our model independent 
constraint, it is clear that 
neutron stars
cannot be ruled out, even if many explicit models can, always
provided the X-ray  phenomena are modeled correctly. 

\section{Caveats}
We have derived a mass-radius relationship for ``neutron stars''
employing the minimal 
conservative constraints enumerated above. In doing so we are recognizing
that there is no empirical knowledge of the properties of nuclear
matter above saturation density of $\sim 0.15 {\rm~fm^{-3}}$. 
In our search for the minimum radius as a function of mass, we have allowed
a constant pressure region to develop above the fiducial density of
$0.1625 {\rm~fm^{-3}}$, the density closest to saturation in the BPS
tables of the low-density \eosp. The minimum radius is increased by $\sim 0.1
$ km if the \eos is merely very soft just above the fiducial density.
These features are permitted by our ignorance above saturation
density, and in this sense provide a conservative estimate of the
minimum
radius as a function of mass.  However, we may be permitted some 
prejudice: If  a low-density
phase transition of any kind were not plausible, then the minimum radius
that we have derived, would be increased.

The central density of the minimum radius star, in either
of the above two cases is about 26 times nuclear density for a 
canonical $1.44\msun$ star. Almost certainly, stars of
such high central density must contain a deconfined quark matter core.
However, these stars, with the constraints that we have imposed, are
gravitationally bound, rather than self-bound, as a strange star would
be. So, if the analyses of these X-ray objects really does
imply an extraordinarily small radius,  that fact  would be consistent with
the star having a quark matter core, the quarks being liberated
from hadrons by the high pressure.  This is in distinction with the 
strange matter hypothesis, according to which the entire star would be
made of self-bound quark matter, a so-far undiscovered state, 
the actual ground state of hadronic matter, if the hypothesis is true.

\section{Comments}
Two or three
properties of pulsars can be measured with great accuracy---the period
of rotation, sometimes the time rate of change of period, and the masses involved in close binaries. The first two are 
directly observed, and the third deduced from measurement of orbital
parameters. With sufficient observation time, these can be determined
accurately, and little doubt surrounds orbital mechanics.
It is possible that no  other properties gained from any other phenomena
will rival these  types of measurements either in accuracy or
in clarity of interpretation. 

The detection of a pulsar with a  rotational
period smaller for its mass than
that obtained as a model independent limit for neutron stars 
would be decisive in distinguishing 
between the neutron star interpretation
of pulsars as compared to an exotic star---a star that is self-bound at
very high equilibrium density
(see Eq.\ (8) in Ref.\
(\cite{glen92:a})). However, nature may never provide a
mechanism for approaching the limiting period, which for a neutron star of
mass $1.44 \msun$ is about $0.3$ ms. 

On the other hand,
the pulsation phenomenon in X-ray
stars may involve a relation between mass and radius
which could also be decisive. However,
the interpretation  is subject to some uncertainty, 
both as to the origin of the pulsations and
most certainly as to the  accuracy of the mass-radius connection,
Eq.\ \ref{kepler}. This relationship
 holds only for classical and for non-rotating relativistic stars.
There {\sl is no formula} for the Kepler frequency
of  a particle orbiting 
{\sl rotating} relativistic stars because of the position dependent
frame-dragging frequency; rather the Kepler
frequency can be determined only as a self-consistency condition
on the solution of Einstein's equations and therefore only for specific 
model assumptions. There is no 
possibility of evading this model dependence; however it becomes weaker
at further distance outside the star. [See Eq. (8) in 
(\cite{glendenning94})].

Therefore, it is of general 
interest to have a model-independent
limit on radius
as a function of mass of neutron stars, 
such as we have provided here, and it is of
particular interest in connection with the oscillations in X-ray 
 brightness of neutron star accreters.\\[2ex]

Acknowledgements:
   I am indebted to I. Bombaci for enlightening comments
   and criticism of an early draft. \doe

\end{document}